\documentstyle[12pt]{article}
\begin{document}
\thispagestyle{empty}
\begin{center}{\Large{Dynamo efficiency in large scale magnetic fields in parity violating torsion theories}}
\end{center}
\vspace{1.0cm}
\begin{center}
{\large By L.C. Garcia de Andrade\footnote{Departamento de
F\'{\i}sica Te\'{o}rica - IF - UERJ - Rua S\~{a}o Francisco Xavier
524, Rio de Janeiro, RJ, Maracan\~{a}, CEP:20550.
e-mail:garcia@dft.if.uerj.br}}
\end{center}
\begin{abstract}
Earlier a generalised dynamo equation in first order torsion [Phys Lett B (2012)] was derived. From this equation it is shown that for $10 kpc$ scale torsion gravity is not able in helping to seed galactic dynamos due to the fact that time is not long enough to take into account structure formation. In this paper we extend this dynamo equation to second-order torsion terms, but unfortunatly situation is even worse and the torsion seems not help dynamo efficiency. Nevertheless in the scale of intergalactic magnetic field of $1Mpc$, the efficiency of the self-induction equation with torsion situation is much better and even in the first-order torsion case, one obtains large scale magnetic fields of $10^{9}yrs$ dynamo efficiency. This seems to be in contrast with recent investigation by Bamba et al [JCAP (2010)] where they obtain from a special  type of torsion theory called teleparallelism [A Einstein, Math Annalen (1922)], a large  scale intergalactic magnetic field of $10^{-9}G$. If this is not a model dependent result there is an apparent contradiction that has to be addressed. It is shown that assumed dynamo efficiency in astrophysical flow without shear a strong seed field of $10^{-11}G$ which is suitable to seed galactic dynamos. \end{abstract}

\newpage
\section{Introduction}

One of the most interesting problems in dynamo theory of large scale fields is the dynamo efficiency \cite{1} which depends essentially on time the dynamo is able to last to amplify magnetic fields up to galactic magnetic amplification the so called galactic dynamo problem \cite{2}. The time the dynamo action lasts is fundamental to decide if the dynamo action is responsable for  cosmic action or not only gravitational compression or Biermann battery. Earlier we have shown \cite{3} that deriving a dynamo equation from a parity violating gravity with torsion the dynamo efficiency is not enough in the first-order torsion equation. Is a general understanding in the cosmology and dynamo communities that dynamo is not enough to explain the intergalactic magnetic fields \cite{4} though is able to seed galactic dynamos \cite{5}. Unfortunatly this seems not to be the case in torsion theories. At scales of $10Kpc$ this is the case however, as shown in this paper galactic dynamo seed, not granted at this scale can reach intergalactic magnetic fields with dynamo efficiency diffusion time of the order of $10^{9}yrs$. Another example from the dynamo equation with shear shows that in the absence of shear a seed field of $10^{-11}G$ on gravitational dynamo or compression of pregalactic structure. The paper is organised as follows: In section 2 we review the derivation of dynamo equation in spacetimes endowed with torsion in analogy to Marklund and Clarkson \cite{6} general relativistic dynamo equation. In section 3  astrophysical consequences are discussed along with generalised dynamo equation with torsion and shear. Section 4 is left for conclusions.
\section{Dynamo mechanism, from parity violating torsion} Parity violating terms $R_{{\mu}{\nu}{\alpha}{\beta}}{\epsilon}^{{\mu}{\nu}{\beta}{\epsilon}}$ along with terms $R_{{\mu\nu\beta\alpha}}F^{{\mu\nu}}F^{\beta\alpha}$ in the Maxwell electrodynamics in spacetimes with torsion leads to the dynamo equation
\begin{equation}
{D}_{\mu}F^{\mu\nu}+\frac{1}{m^{2}}D_{\mu}[4{R^{{\mu\nu}}}_{\rho\sigma}F^{\rho\sigma}]=0 \label{1}
\end{equation}
the corrected formula is
\begin{equation}
{\partial}_{\mu}F^{\mu\nu}+T_{\mu}F^{{\mu\nu}}=0
\label{2}
\end{equation}
where we have used $D_{\mu}={\partial}_{\mu}+T_{\mu}$ where $T_{\mu}$ $({\mu}=0,1,2,3)$ , $D_{\mu}$ is the covariant derivative in Riemann-Cartan spacetime. Thus the last expression was obtained taking the torsion field up to first order. In equation (\ref{2}) we have also considered the Bianchi identity
\begin{equation}
D_{\mu}[{R^{{\mu\nu}}}_{\rho\sigma}]=0
\label{3}
\end{equation}
From generalised vector Maxwell equations with torsion one obtains the dynamo equation
\begin{equation}
{\partial}_{t}\textbf{B}-{\nabla}{\times}[\textbf{V}{\times}\textbf{B}]=0
\label{4}
\end{equation}
Here $\eta$ is the diffusion or the electric resistivity of the plasma flow. In the  next section we show the main results of this communication.
\section{Dynamo efficiency on the amplification of magnetic fields}
In the structure formation region of the universe we consider that in galaxies the diffusion constant is important and as in galactic disks as $\eta\sim{10^{26}cm^{2}s^{-1}}$. As noticed before if one computes the time necessary for the dynamo action lasts until the galaxy formation be reach, yields \cite{1}
\begin{equation}
{\tau}_{d}\sim{\frac{{B}}{{\eta}{\Delta}B}}
\label{5}
\end{equation}
Substitution of the generalised torsion term in the dynamo equation to compare its effects with general relativistic ones
\begin{equation}
{\tau}_{d}\sim{\frac{{B}}{{\eta}({\nabla}\textbf{T})B}}
\label{6}
\end{equation}
where $T$ here is the actual terrestrial torsion
given in Laemmerzahl \cite{7} given by $T\sim{10^{-17}cm^{-1}}$ or $10^{-31}GeV$. Therefore this equation represents
\begin{equation}
{\tau}_{d}\sim{\frac{L}{{\eta}T}} \label{7}
\end{equation}
note that to $10kpc$ scale we obtain ${\tau}_{d}\sim{10^{6}yrs}$ which is still far from structure formation and torsion cannot generate dynamo efficiency, however if the scale of magnetic fields is larger as $1Mpc$ then the same torsion formula above yields $10^{9}yrs$ which is enough to yield the galactic dynamo seeds and dynamo efficiency. It is easy to find that for the galactic dynamo to be effective in second order torsion one reaches only $10^{-7}yrs$ which is still far beyond to grant the dynamo efficiency.
\section{Large magnetic fields from dynamo equation without shear}
Gnedin et al \cite{8} have argued that $10^{-20}G$ of seed fields by gravitational compression could not be useful to seed galactic dynamos. In this section we show that gravitational dynamos can be obtained from Lorentz violation using scalar torsion given by $S_{ijk}=S_{0}{\epsilon}_{ijk}$. From Widrow \cite{1} we may consider the generalization of dynamo equation with torsion to include an important physical feature of cosmological flows which is the shear. This equation can easily derived from
\begin{equation}
\frac{dB_{i}}{dt}=\frac{2}{3}\frac{B_{i}}{\rho}\frac{d{\rho}}{dt}+B^{j}({\sigma}_{ij}+S_{0}{\epsilon}_{ijk}V^{k})
\label{8}
\end{equation}
where the shear tensor is given by
\begin{equation}
{\sigma}_{ij}=[{\partial}_{j}V_{i}-\frac{1}{3}{\nabla}.\textbf{V}]
\label{9}
\end{equation}
Torsion effects can be eliminated by using the expression
\begin{equation}
{\sigma}_{ij}=-{\epsilon}_{ijk}S_{0}V^{k}
\label{10}
\end{equation}
A particular example to galactic disk can be given now taken as the only surviving components $V=V_{y}$ and $B=B^{z}$. These substitutions on dynamo equation in the absence of shear yields
\begin{equation}
\frac{1}{B}\frac{dB}{dt}=\frac{2}{3}\frac{1}{\rho}\frac{d{\rho}}{dt}-S_{0}V
\label{11}
\end{equation}
this equation can be transformed to a more suitable integrable form
\begin{equation}
\frac{d\frac{lnB}{{\rho}^{\frac{2}{3}}}}{dt}=-S_{0}V
\label{12}
\end{equation}
by integrating this equation one obtains
\begin{equation}
B\sim{{{\rho}^{\frac{2}{3}}}S_{0}L}
\label{13}
\end{equation}
substitution of dates $L\sim{1kpc}$, ${\rho}\sim{10^{-24}g.cm^{-3}}$ and torsion value above one obtains the a cosmological magnetic field seed of $10^{-11}Gauss$ which is a seed field strong enough to seed galactic dynamos.

\section{Conclusions}
The issue of origin of cosmic magnetic field has been a matter of  controversy specially in terms of torsion. Recently K Bamba et al \cite{9} has shown that by using a special type of gravity theory of torsion earlier called by Einstein by the name teleparallelism. In this paper we show that by making use of a dynamo equation with torsion, a non-teleparallel theory could actually have an efficient dynamo able to seed galactic magnetic fields as long we use large scales as $1Mpc$. It is interesting to note that from the dynamo equation in Riemann-Cartan spacetime torsion is useless in highly conductive phases since it is coupled to the magnetic diffusion. This situation is similar to the mean field dynamos where viscosity which would be useful only in the case of stationary dynamos. Seed fields of the order of $10^{-11}G$ are obtained.

\section{Acknowledgements} I would like to express my gratitude to Rajeev K Jain and Professor J Yokohama for their many enlightning discussions on the problem of magnetogenesis and dynamo efficiency both at Institute d'Astrophysique at Paris meeting on Primordial Universe after Planck mission held in december 2014. Financial support from University of State of
Rio de Janeiro (UERJ) are grateful acknowledged.

\end{document}